# Measurement of $^{37}$Ar to support technology for On-site Inspection under the Comprehensive Nuclear-Test-Ban Treaty

C.E. Aalseth, A.R. Day, D.A. Haas, E.W. Hoppe, B.J. Hyronimus, M.E. Keillor, E.K. Mace, J.L. Orrell,[*] A. Seifert, V.T. Woods

*Pacific Northwest National Laboratory, 902 Battelle Boulevard., Richland, WA 99352, USA*

**Abstract**

On-Site Inspection (OSI) is a key component of the verification regime for the Comprehensive Nuclear-Test-Ban Treaty (CTBT). Measurements of radionuclide isotopes created by an underground nuclear explosion are a valuable signature of a Treaty violation. Argon-37 is produced from neutron interaction with calcium in soil, $^{40}$Ca(n,$\alpha$)$^{37}$Ar. For OSI, the 35-day half-life of $^{37}$Ar provides both high specific activity and sufficient time for completion of an inspection before decay limits sensitivity. This paper presents a low-background internal-source gas proportional counter with an $^{37}$Ar measurement sensitivity level equivalent to 45.1 mBq/SCM in whole air.

## 1. Introduction

Nuclear reaction products are the ultimate confirmatory signatures an explosion was nuclear in nature, with chemically inert noble gases being the most likely to escape into the environment. Radioxenon collection and analysis is a standard technique used to monitor for evidence of nuclear detonations. More than fifty days past an underground nuclear detonation, the $^{37}$Ar signature should be stronger than the radioxenon signature. This prediction was verified during Project Gasbuggy, a 27-kt underground test. Gas samples from a well drilled near the cavity showed the initial activity was on the order of 75-200 MBq per standard cubic meter (MBq/SCM) air [1].

The measurement of $^{37}$Ar is well understood but challenging, typically employing low-background proportional counters where the $^{37}$Ar gas is included in 90% argon / 10% methane (P10) counting gas. The 35-day half-life of $^{37}$Ar makes it amenable to collection, recovery, and measurement after a nuclear detonation. The challenge resides in measuring the residual X-rays and/or Auger electrons from the ground-to-ground state electron capture decay of $^{37}$Ar to $^{37}$Cl. The atomic shell cascade process results in an energy peak signature in a proportional counter at 2.822 keV [2]. The proportional counter must therefore have both a low energy threshold (< 1 keV) as well as a low intrinsic radioactive background.

In the context of On-Site Inspection (OSI) under the auspices of the Comprehensive Nuclear-Test-Ban Treaty (CTBT), the MARDS system [3] was developed to separate argon from whole air and measure for $^{37}$Ar activity. A lowest background rate of 0.16 counts/second under the $^{37}$Ar peak was reported for the MARDS proportional counter [3]. This paper presents initial development of an ultra-low-background proportional counter for measurement of $^{37}$Ar relevant to CTBT OSI.

_________

[*] Corresponding author. Tel.: +1-509-376-4361; fax: +1-509-376-3868; e-mail: john.orrell@pnl.gov.

## 2. Production and transfer of radioargon

Two methods were considered for generating an $^{37}$Ar sample. The first method is the $^{40}$Ca(n,α)$^{37}$Ar process where neutrons are supplied by a nuclear reactor. This is the same (n,α) reaction process that produces $^{37}$Ar from calcium in rock and soil surrounding an underground nuclear detonation. The second method is irradiation of natural argon with reactor neutrons. The second method was selected based upon the authors' familiarity with irradiation and handling of noble gases.

A 1 cm$^3$ volume of commercial high purity natural argon was irradiated with neutrons at an in-core position in the University of Texas at Austin TRIGA reactor. Figure 1 shows an example calculation of the radioargon isotope's activity levels versus time after a 3600 second irradiation. For the sample produced, the actual irradiation time was ~5400 seconds.

The $^{41}$Ar 1294-keV γ-ray (99.1% intensity) provides a convenient means for estimating the $^{37}$Ar activity. Approximately 1 day after irradiation, the argon was transferred into a stainless steel 50 cm$^3$ shipping-container and counted at a standoff distance using a high purity germanium (HPGe) γ-ray spectrometer with geometric detection efficiency measured by an $^{152}$Eu standard. This determined the $^{41}$Ar activity was 12.40±0.17 MBq at the end of the irradiation period. The uncertainty (1.4%) is due primarily to the counting statistics of the $^{41}$Ar 1294-keV γ-ray peak. The $^{37}$Ar activity is determined through a calculated ratio of the production cross sections of $^{40}$Ar(n,α)$^{41}$Ar to $^{36}$Ar(n,γ)$^{37}$Ar. This ratio is uncertain at the 10% level due to the knowledge of the neutron spectrum at the location of the irradiation vessel in the reactor, resulting in a calculated initial $^{37}$Ar activity of 1.31±0.13 kBq. After transport to PNNL, a 1.34±0.18 Bq aliquot of gas was loaded from the shipping-container into an ultra-low-background proportional counter (ULBPC). The ULBPC was then back-filled to 1516 Torr with P10 counting gas.

Prior to loading the aliquot of gas into the ULBPC (~7 days after irradiation), the shipping container was counted for ~4 days on a low-background HPGe detector. No γ-ray peaks were found beyond those attributed to ubiquitous environmental radioisotopes.

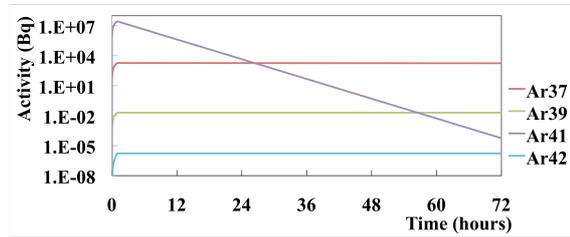

Fig. 1. Calculated activity of radioargon isotopes from a 3600 second in-core neutron irradiation of 1 cm$^3$ of natural argon gas.

## 3. Proportional counter measurement and analysis

The detector used is the product of an effort to produce a low-background, physically robust gas proportional counter for applications like radon emanation measurements, groundwater tritium, and $^{37}$Ar [4]. The ultra-low-background proportional counter (ULBPC) was designed to handle a wider range of gas volumes, have higher efficiency, and be easier to assemble than current ultra-low background solutions. The materials chosen for the detector were specifically selected for their proven radiopurity. High-purity electroformed copper offers excellent radiopurity along with good electrical, thermal, and vacuum properties. The ULBPC's body is an electroformed copper cylinder roughly 9" in length having a ~0.1 L internal volume. The plastic CPTFE, which forms the gas seal and provides high voltage insulation, was chosen for its very low gas permeability and reasonable measured background level. The ULBPC's anode wire is 0.001-inch-diameter niobium, a radiopure material used successfully in the IGEX experiment [5].

The ULBPC was operated in an aboveground lead shield consisting of 2", 4", and 8" thick walls (top, sides, and bottom, respectively). Two 2" polyvinyl toluene (PVT) cosmic ray veto panels were placed above and below the lead shield, each having 33"×14" are 56"×19" footprints, respectively.

A CAEN A1833P supplied high voltage to the veto panel photomultiplier tubes (PMTs) and the ULBPC. The bottom veto panel's single Saint-Gobain 200-5888 PMT was operated at 945 V. The top veto panel had two Saint-Gobain 200-5887 PMTs operated at 850 V and 890 V. PMT high voltage was chosen such that an oscilloscope showed the signal pulse output from each PMT were all roughly matched in gain. Two Canberra 2005 preamplifiers (low-gain setting) amplified the PMT's signal output. The top veto panel's PMT outputs were combined (BNC T-connector) prior to entering the preamplifier, and then followed by a 50-Ohm signal attenuator. The ULBPC was operated at +2100 V with a Canberra 2006 preamplifier set for high-gain. The ULBPC and veto panels' signals were acquired using a XIA PIXIE-4 waveform-digitizing card. Minimum digital gains on the PIXIE-4 card input channels resulted in a broad range of energy deposition sensitivity to cosmic rays in the veto panels (Fig. 2) and an energy range of 0-20 keV in the ULBPC (Fig. 3). Pulse waveforms from the three input channels were digitized and stored for offline analysis whenever a pulse trigger was detected on the ULBPC channel.

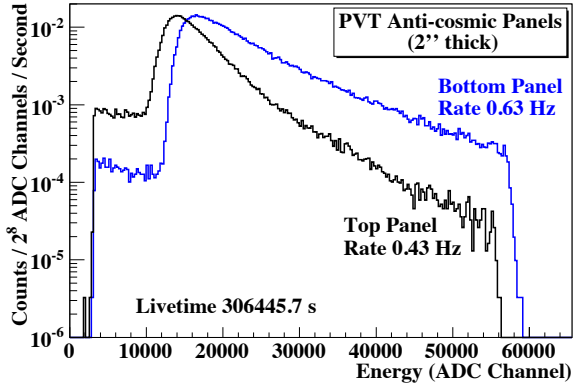

Fig. 2. Un-scaled energy spectra from the cosmic ray veto panels when triggered in coincidence with the proportional counter.

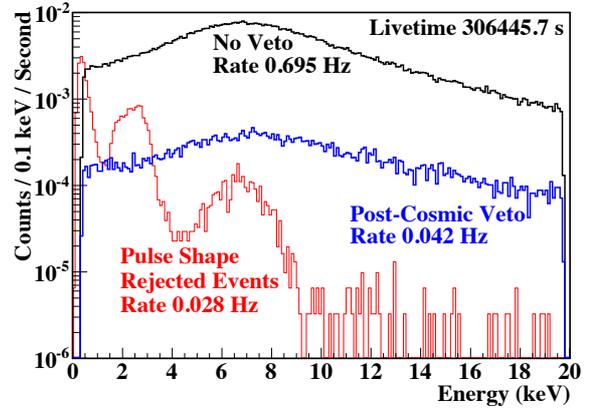

Fig. 3. Proportional counter background spectra before and after application of a cosmic ray veto. Also shown is the spectrum of electronic noise events removed via a pulse shape risetime cut.

For energy calibration, $\gamma$-rays from ~80 $\mu$Ci of $^{241}$Am interact in the copper walls of the ULBPC resulting in emission of K-shell X-rays. The five highest intensity Cu K-shell X-rays [6] have an intensity weighted mean energy of 8.133 keV. To obtain a two point linear energy scale fit, the $^{37}$Ar peak at 2.822 keV [2] is also used. Over a range of high voltage values from +1500 V to +2100 V, the full width half maximum energy resolution of the Cu K-shell X-ray peak varied by less than 1%; the +2100 V bias voltage for the ULBPC was selected to provide gas gain for detection of lower energy events.

Prior to the $^{37}$Ar load, the background of the ULBPC was studied (Fig. 3). A population of fast-rising electronic-noise pulses was identified and removed by a pulse shape analysis requiring the 10%-90% risetime of the pulse be less than 798 ns. The remaining "No Veto" events in Fig. 3 compose the full ULBPC background. Requiring neither veto panel has fired reduces the ULBPC background by ~16.5 times. The broad peak of the ULBPC background spectra seem consist with an estimation of ~9.5 keV deposited by minimum ionizing muons passing through the count gas volume. The similar shape of the two ULBPC background spectra in Fig. 3 is indicative of a remaining predominance of cosmic ray induced background events as well as insensitivity to ubiquitous $\gamma$-ray backgrounds. The former conclusion is expected considering the limited coverage provided by the two veto panels.

Figure 4 shows the ULBPC's spectra. Before the $^{37}$Ar sample load, the peak at 2.8 keV is absent. The first and the last data collection runs show the decay of the intensity of the peak located at 2.8 keV. The inset to Fig. 4 shows the continuum-subtracted rate of events in the 2.8 keV peak as a function of time since the reactor irradiation. Despite the gap in measurements from days 43 to 55, an exponential fit estimates a half-life of 35.6±0.3 days, in reasonable agreement with the standardized [6] 35.04±0.04 day half-life for $^{37}$Ar. Using the fit results for the half-life decay value, an estimate of 1.36±0.02 counts/seconds would have been found in the $^{37}$Ar peak at the time of the ULBPC load. Accounting for the 90.2% branch to the K-shell for the $^{37}$Ar decay one arrives at an activity estimate of 1.50±0.02 Bq based on the count rate measured beneath the 2.8 keV peak.

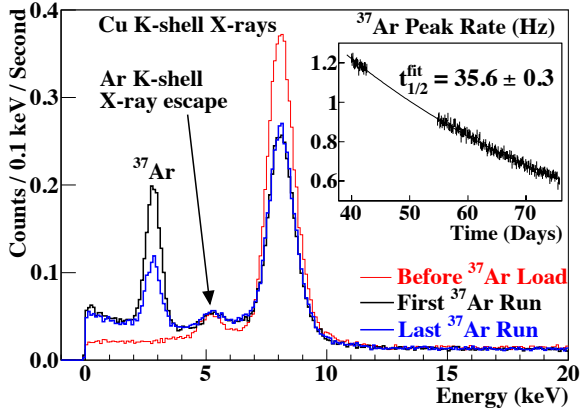

Fig. 4. Proportional counter energy spectra. When the counter was loaded with the $^{37}$Ar sample, the $^{241}$Am calibration source was re-oriented resulting in a change in the intensity of the Cu K-shell X-ray peak. Inset shows the half-life decay of the $^{37}$Ar peak intensity.

## 4. Discussion

In relation to CTBT/OSI, the quantity of interest is the sensitivity level at which a detector can measure the presence of $^{37}$Ar in whole air. Once the argon gas fraction is separated from the initial whole air volume, the proportional counter is in essence performing an $^{37}$Ar specific activity measurement. Applying the methods of Currie [7] to the presently described ULBPC, the sensitivity for a hypothetical measurement for the presence of $^{37}$Ar in argon extracted from whole air is determined as follows.

First, a region of interest (ROI) is defined in the energy spectrum of the ULBPC from 2.0 – 3.6 keV. This ROI corresponds to ±2.6 times the $^{37}$Ar peak's Gaussian width ($\sigma$ = 0.3 keV) resulting in ~99% coverage of the events in a $^{37}$Ar peak. The integral rate of background events after application of the risetime cut and cosmic-ray veto (see Fig. 2) is 0.003 counts/second. A measurement count time of 10 hours is assumed resulting in a mean number of background counts $\mu_B$ = 108 in the ROI. False positive and false negative errors are set at 10% (i.e. $\alpha$ = 0.10 and $\beta$ = 0.10). As defined by Currie [7], the $L_C$ = 121 counts in the ROI during the 10-hour count duration. The corresponding $L_D$ = 136.4 counts in the

ROI during the 10-hour count duration. Thus to satisfy the desired false positive and false negative rates (i.e. $\alpha$ and $\beta$ values of 10%), $L_D - \mu_B$ = 28.4 counts above the mean background rate are required in the ROI during the 10-hour count duration to conclude $^{37}$Ar has been detected. To scale this number to a measurement of $^{37}$Ar specific activity in whole air, a constant is defined as

$$C_1 = \frac{1}{\varepsilon}\left(\frac{f_{Ar}}{V_{pc}}\right)\left(\frac{1}{R_{K\text{-shell}}T_{count}}\right)\left[\frac{10^3 mBq}{Bq}\frac{10^3 L}{SCM}\right]$$

with an assumed proportional counter measurement efficiency of $\varepsilon$ = 90% for the K-shell branch following $^{37}$Ar decay, the fraction of argon gas in whole air $f_{Ar}$ = 0.93%, the maximum volume of argon gas in a 2 atmosphere load of the proportional counter $V_{pc}$ = 0.2 L, the branching ratio to the K-shell after $^{37}$Ar decay $R_{K\text{-shell}}$ = 90.2% [2], and a measurement count time of $T_{count}$ = 36000 seconds. Thus, the whole-air specific activity (i.e. MDC = minimum detectable concentration) for $^{37}$Ar corresponding to each of Currie's detection limits is:

- ($L_C - \mu_B$) $C_1$ = 20.7 mBq/SCM
- ($L_D - \mu_B$) $C_1$ = 45.1 mBq/SCM

This result is interesting because sub-surface measurements [8] have shown naturally occurring background levels of $^{37}$Ar are present at the 10-100 mBq/SCM level. Finally, note this calculation assumes argon gas is extracted from an unlimited quantity of whole air. In other words, as much whole air is processed as is needed to produce the 0.2 L of argon gas needed to fill the proportional counter. If argon gas extraction efficiency from whole air is estimated at 40%, the above measurement with the quoted specific activity sensitivity could be made from a 54 L sample of whole air.

## Acknowledgments

The authors thank Steven R. Biegalski, University of Texas at Austin, for preparing the $^{37}$Ar samples. This research was supported by the Laboratory Directed Research and Development Program at the Pacific Northwest National Laboratory (PNNL) operated by Battelle for the U.S. Department of Energy under Contract DE-AC05-76RL01830.